\documentclass{iopart}
\usepackage{epsfig,graphicx,graphics}
\usepackage{flafter}
\usepackage{color}
\usepackage{bm}
\usepackage{amssymb}
\usepackage{epstopdf}
\usepackage{amstext}
\usepackage{amsthm}
\usepackage{amsfonts}
\usepackage{latexsym}
\usepackage{array}

\begin{document}

\title{Coherent Quantum Phase Slip in two-component bosonic Atomtronic Circuits} 

\author{A. Gallem\'{\i} $^{1,2}$, A. Mu\~noz Mateo $^1$, R. Mayol $^{1,2}$ and M. Guilleumas $^{1,2}$}
\address{$^1$ Departament d'Estructura i Constituents de la Mat\`{e}ria, Facultat de F\'{\i}sica, Universitat de Barcelona, E--08028 Barcelona, Spain}
\address{$^2$ Institut de Nanoci\`encia i Nanotecnologia de la Universitat de Barcelona, IN$\,^2$UB, E--08028 Barcelona, Spain}

\date{\today}

\begin{abstract}
Coherent Quantum Phase Slip consists in the coherent transfer of 
vortices in superfluids. We investigate this phenomenon in two 
miscible coherently coupled components of a spinor Bose gas 
confined in a toroidal trap. After imprinting different vortex 
states on each component, we demonstrate that during the whole 
dynamics the system remains in a linear superposition of two 
current states in spite of the 
non-linearity and can be mapped onto a linear Josephson problem. 
We propose this system as a good candidate for the realization 
of a Mooij-Harmans qubit and remark its feasibility for implementation 
in current experiments with $^{87}\mbox{Rb}$, since we have used values 
for the physical parameters currently available in laboratories. 
\end{abstract}


\section{Introduction}

The possibility to set up experiments devoted to the test of quantum phenomena 
has developed a singly peerless increasing interest since the discovery of 
superconductivity. At an early stage, this research led to the realization of Superconducting Quantum 
Interference Devices (SQUIDs) \cite{Jaklevic1964,Jaklevic1965}. Three decades later, 
the achievement of 
ultracold degenerate quantum gases, like Bose-Einstein condensates (BECs)  
\cite{Anderson1995,Davis1995}, opened up new opportunities to test quantum interference 
phenomena, and their implementation in Atomtronic Quantum Interference Devices 
(AQUIDs) \cite{Seaman2007,Ryu2007,Moulder2012,Wright2013,Ryu2013,Eckel2014}, 
the atomic analogue of SQUIDs. This duality between superfluid atomic gases and 
superconductors has stood out 
both systems as good supports for quantum simulation. However, the 
tunability of the interaction and the versatility of atoms in simulating 
both bosonic and fermionic systems, provide a more promising 
perspective for the implementation of AQUIDs in future technological applications. 

Josephson junctions play a key role in the physics of quantum interference devices. 
They are constituted by two quantum systems connected by a weak link, and can 
be classified in two categories (short and long), owing to the different nature 
of the coupling. In a short 
Josephson junction, the coherent transfer of physical quantities occurs 
through a single point, the Josephson link \cite{Albiez2005,Levy2007}. 
When the junction is long, the coupling occurs locally at each point of the 
connection. In particular, two spin components of a 
condensate coupled by a Raman laser operate as a long Josephson junction, which is  
referred in the literature as internal Josephson effect \cite{Sols1999}, and 
obeys the Josephson equations with a coupling proportional 
to the overlap between condensates \cite{Williams1999b}. 
These techniques, by selecting an appropriate spatial dependence of the 
coupling, led to the first observation of vortices in 
BECs \cite{Matthews1999,Williams1999a}.

Anderson \cite{Anderson1966} discussed the role of the phase of 
the order parameter in superfluids, which motivated the study of 
phase slips in superconductivity \cite{Arutyunov2008}, liquid Helium 
\cite{Varoquaux2015}, and BECs \cite{Wright2013,Eckel2014}.
A phase slip event is a sudden change of the phase in $2\pi$ due to the 
motion of quantized vortices through a superfluid. This phenomenon is 
associated to dissipation, as pointed out by Langer and Ambegaokar 
\cite{Langer1967}. Ultracold atoms, as superfluids, can also exhibit 
phase slips, by winding the phase through solitonic states 
\cite{Piazza2009,Piazza2013,Abad2011a,Abad2015,MunozMateo2015} 
and they are able to generate quantum superpositions of 
macroscopic flows. The literature includes several proposals to engineer superpositions of 
flow 
states in 1D quantum gases in continuous rings \cite{Scheke2011,Halkyard2010,Aghamalyan2013} and discrete 
rings 
\cite{Aghamalyan2015,Hallwood2010,Hallwood2011,Gallemi2015}.

Coherent Quantum Phase Slip (CQPS) is an effect recently discovered in 
superconducting systems containing loops \cite{Astafiev2012}. It is the 
dual phenomenon of the Josephson effect, which is a coherent transport of 
particles between two superfluids, but, in contrast, 
CQPS is defined as the coherent transfer of 
vortices through the Josephson link. In CQPS, 
the stationary states corresponding to flux 
quanta states become coupled, such that one can continuously 
change the flux quanta of the system. 
The Mooij-Harmans qubit \cite{Mooij2005,Mooij2006}, which 
consists of a superconducting loop with a weak nanowire, was predicted to be 
able to manifest CQPS between two current states. 
The proposal was made a reality in the experiment of 
Ref. \cite{Astafiev2012}, which led to the first 
experimental observation of CQPS.

In this work, in order 
to implement CQPS, we propose the realization of an atomic analogue of the Mooij-Harmans 
qubit by means of a spinor condensate with two relevant internal degrees of freedom or 
spin states. The two components are coupled by phase (spin exchange) and density (contact 
interaction), and both occupy the same space region, since the interspecies density 
repulsion is small enough to keep the system in the miscible phase. Therefore, this 
overlap allows the coupling to occur locally, point to point, in the whole bulk of the 
condensate (long Josephson junction). With the aim to engineer a qubit we will select 
vortex states as the basis of an effective two-level system that is able to perform 
qubit operations \cite{Lim2014}. The coherent coupling transfers vortices between 
both components in the absence of population imbalance, and the non-linear system 
exhibits Rabi oscillations. We consider mixtures confined in ring geometries, 
where persistent currents are metastable states and phase slips provide the 
mechanism for the system to exchange winding numbers between components. 
All these properties stand out the system as a promising tool for atomtronic circuits, 
and in particular, for the simulation of CQPS. 

The paper is organized as follows. In section \ref{model}, we present 
the mean field model we have used to study the system of two miscible 
coherently-coupled condensates. Section \ref{dynreg} characterizes the 
different dynamical regimes that the system exhibits as a function of 
the coherent coupling and interaction. In section \ref{cqps} we discuss 
the regime where the system shows CQPS and propose an analytical model 
that accurately reproduces our numerical results. Section \ref{Other} 
is devoted to the other dynamical regimes of the phase diagram, and 
finally, section \ref{conclusion} summarizes our work and provides 
future perspectives.

\section{Theoretical framework}
\label{model}

To describe the system of two coherently coupled Bose-Einstein condensates in the mean field 
regime, we will use the Gross-Pitaevskii equation (GPE) for the wavefunctions $\Psi_\uparrow$ and $\Psi_\downarrow$: 
\begin{eqnarray}
 i\hbar\frac{\partial}{\partial t}\Psi_\uparrow&=\mathcal{H}_0\Psi_\uparrow+
 g_{\uparrow\uparrow}|\Psi_\uparrow|^2\Psi_\uparrow+g_{\uparrow\downarrow}|\Psi_\downarrow|^2\Psi_\uparrow+\frac{\hbar\Omega}{2}\Psi_\downarrow\nonumber\\
 i\hbar\frac{\partial}{\partial t}\Psi_\downarrow&=\mathcal{H}_0\Psi_\downarrow+
 g_{\downarrow\downarrow}|\Psi_\downarrow|^2\Psi_\downarrow+g_{\uparrow\downarrow}|\Psi_\uparrow|^2\Psi_\downarrow+\frac{\hbar\Omega}{2}\Psi_\uparrow\,,
 \label{tdgpe}
\end{eqnarray}
where $\mathcal{H}_0=-\hbar^2/2m \,\vec{\nabla}^2+V_{\rm trap}$, 
$g_i=4\pi\hbar^2 a_i/m$, with $i=\uparrow\uparrow, \downarrow\downarrow, \uparrow\downarrow$,  
is the interaction strength, $a_i$ is the scattering length, and $\Omega$ 
is the coherent Raman coupling that forces the two components to share the
chemical potential $\mu$. 
We will consider multiply-connected geometries by using a 
toroidal trap of radius $R$, $V_{\rm{trap}}=1/2\,m 
\,\omega_\rho^2((\rho-R)^2+\lambda^2z^2)$, with 
angular frequency $\omega_\rho$, and aspect ratio $\lambda\gg1$.
For quasi-2D systems the interaction strength takes the value
$g^{\rm 2D}_i=g_i\sqrt{8\pi\lambda}$.

The time-independent solutions $\Psi(\vec r,t)=\phi(\vec r)\exp(-i\mu t/\hbar)$ 
of Eq. (\ref{tdgpe}) can be found by solving:
\begin{eqnarray}
 \mu\phi_\uparrow&=\mathcal{H}_0\phi_\uparrow+
 4\pi a_{\uparrow\uparrow}|\phi_\uparrow|^2\phi_\uparrow+4\pi a_{\uparrow\downarrow}|\phi_\downarrow|^2\phi_\uparrow+\frac{\Omega}{2}\phi_\downarrow\nonumber\\
 \mu\phi_\downarrow&=\mathcal{H}_0\phi_\downarrow+
 4\pi a_{\downarrow\downarrow}|\phi_\downarrow|^2\phi_\downarrow+4\pi a_{\uparrow\downarrow}|\phi_\uparrow|^2\phi_\downarrow+\frac{\Omega}{2}\phi_\uparrow\,,
\end{eqnarray}
where all the quantities are written in terms of the harmonic oscillator units, by using $\hbar\,\omega_\rho$ and $\sqrt{\hbar/m\,\omega_\rho}$ 
as energy and length units, respectively. In our simulation we also consider $a_{\uparrow\uparrow}=a_{\downarrow\downarrow}=a \gtrsim a_{\uparrow\downarrow}$ 
to ensure miscibility. The total number of particles $N$ is fixed, in such a way that $\sum_i\int \phi_i^*(\vec{r}) \phi_i (\vec{r}) d\vec{r}=N$, 
though particles of both components can exchange their spin by 
virtue of the coherent Raman coupling.

Analytical expressions can be obtained within the Thomas-Fermi approach in 
order to study the ground state properties of the mixture in the regime of 
large interactions. In the case without Raman  
coupling \cite{Ho1996,Riboli2002,Polo2015}, the Thomas-Fermi wavefunction is the same 
for both components: 
$\psi^{\rm{TF}}=\sqrt{(\mu-V_{\rm trap})/4\pi(a+a_{\uparrow \downarrow})}$. 
Such expression will be also useful in the presence of Raman coupling, where 
the ground state satisfies $\psi_\uparrow=-\psi_\downarrow$ 
\cite{Brand2010} for $\Omega>0$, since a
phase difference of $\pi$ between both components minimizes the mean field 
energy. Therefore, the Raman coupling $\Omega$ becomes 
a simple shift of the chemical potential and the Thomas-Fermi wavefunctions can
be written as
\begin{equation}
 \psi_\uparrow^{\rm TF}=-\psi_\downarrow^{\rm TF}=\sqrt{\frac{\mu_{\rm eff}-V_{\rm trap}}{4\pi(a+a_{\uparrow \downarrow})}}\,,
  \label{TF}
\end{equation}
where $\mu_{\rm eff}=\mu+\Omega/2$ will play the role of an effective chemical potential 
for the coupled system. 

We will also explore the effect of vortices on stationary states. Their associated angular 
momentum per particle is $\hbar q$ for each component, where $q$ is the winding number or 
charge of the vortex, when it is centred. This is no longer true for off-centred vortices. 
Nevertheless, one can give an expression for the dependence of the angular momentum as a 
function of the position $r_i$ of the off-centred vortex, by following the procedure developed 
in Refs. \cite{Guilleumas2001,Pethick2002} for the case of a condensate in a harmonic trap 
in the Thomas-Fermi limit, since in this regime, the effect of the vortex on the density 
profile can be neglected. We have derived the expression for the contribution on the angular 
momentum per particle of a vortex in the density region of a 2D ring: 

\begin{equation}
\frac{L_z^i}{N\hbar q} = 
\frac{3}{32}\frac{\delta}{R}(1-\eta^2)^2-\frac{\eta}{8}(3-\eta^2)+1\,,
\label{off-centred-vortex}
\end{equation}
where $\eta=(r_i-R)/\delta$ and 
$R\pm\delta$ are the external and internal Thomas-Fermi radius, respectively, 
and  $\delta=\sqrt{2 \mu}=[3(a+a_{\uparrow\downarrow})N/2R]^{1/3}$ is the 
half-width of the torus 
in harmonic oscillator units. 

\section{Dynamical regimes of two coupled condensates in a ring}
\label{dynreg}

\begin{figure}[b!]
\centering
\includegraphics[width=0.8\linewidth, clip=true]{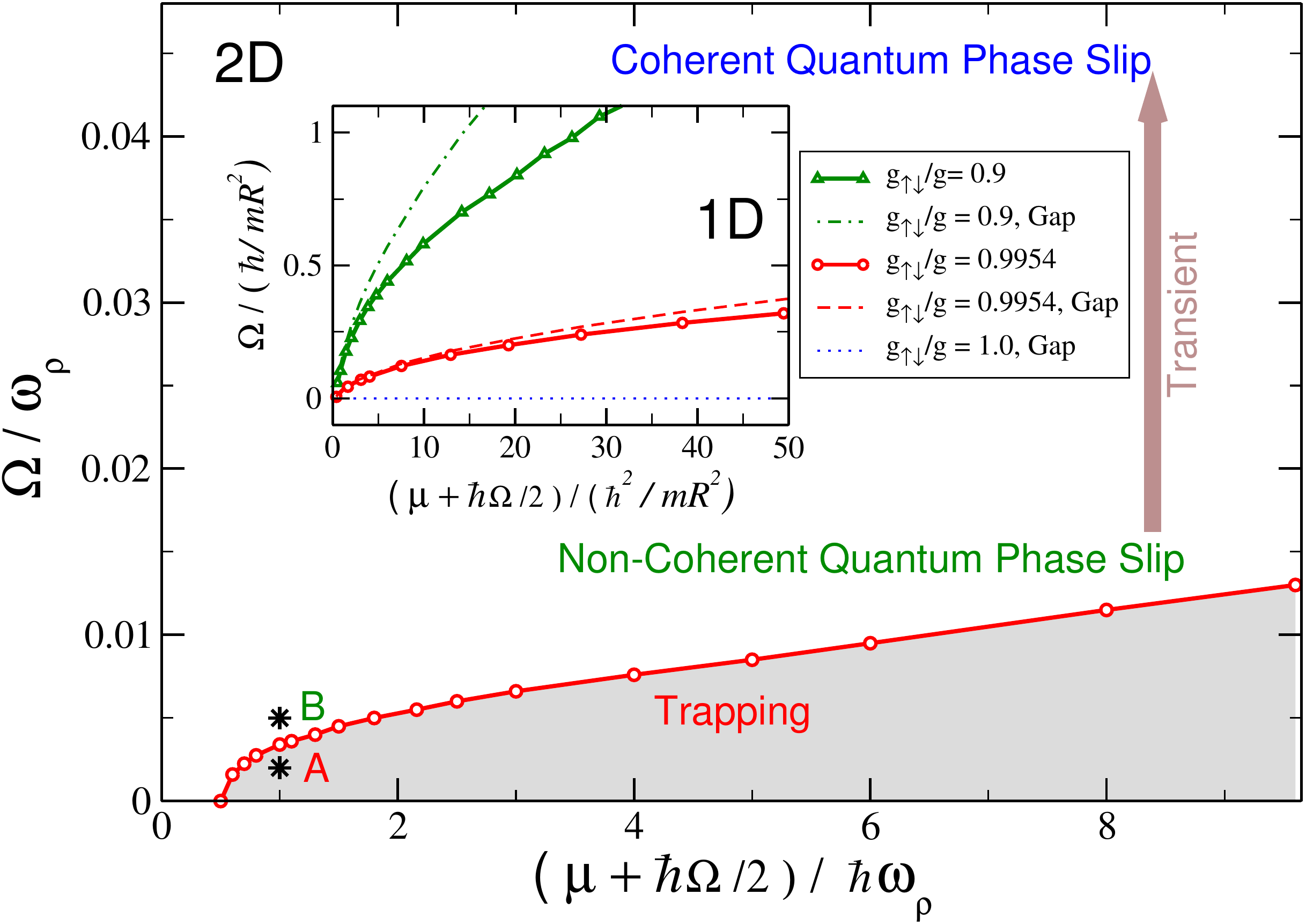}
\caption{Phase diagram containing the different dynamical regimes as a 
function of the effective chemical potential $\mu_{\rm eff}=\mu+\hbar\Omega/2$ 
and the Raman coupling $\Omega$, for $g_{\uparrow\downarrow}/g=0.9954$, radius 
$R=7.5\mu$m, evolving from the initial state $|q_1=1$,$q_2=0\rangle$ in 2D-GPE (\ref{tdgpe}). 
Solid red line draws the boundary $\Omega_c$ between the Trapping regime and the regimes where phase slip exists. 
Above $\Omega_c$ there is a continuous transition from a Non-Coherent Quantum Phase Slip 
regime to a Coherent Quantum Phase Slip regime. The inset 
compares our numerical results for $\Omega_c$ in 1D systems (solid curves with 
open symbols) with the analytical expression Eq. (\ref{gap}) (dotted and dashed 
lines) for the energy gap associated to the excitation of spin modes at 
different values of $g_{\uparrow\downarrow}/g$. The labelled points A, B  
correspond to particular cases addressed in later sections.}
\label{Fig1}
\end{figure}

We investigate the transfer of vortices between two coherently coupled 
Bose-Einstein condensates by solving numerically the time dependent GPE 
(\ref{tdgpe}). To this aim, we have selected typical experimental values of the physical 
parameters. We first compute the ground state of a two-component $^{87}\mbox{Rb}$ 
spinor condensate with intraspecies scattering length 
$a=101.41\,a_B$ and $a_{\uparrow\downarrow}=100.94\,a_B$, 
where $a_B$ is the Bohr radius, corresponding to the hyperfine states 
$|F=1,m_F=0\rangle$ and $|F=1,m_F=-1\rangle$. 
The system is confined in a toroidal trap, with frequency 
$\omega_\rho=2\pi\times200\,\mbox{Hz}$, aspect ratio $\lambda=4$, and 
radius $R=7.5 \,\mu m$. Afterwards, persistent currents are induced in each 
component, with different winding numbers $q_\uparrow$ and $q_\downarrow$, 
by imprinting proper phases, i.e. $\Psi_i\rightarrow \Psi_i\times \exp(i q_i \theta)$, 
and  the system, whose state will be described 
by the pair $|q_\uparrow$,$q_\downarrow\rangle$, is let to evolve. 

In Fig. \ref{Fig1} we show the dynamical regimes of the system, for the initial 
state $|q_1=1$,$\,q_2=0\rangle$, obtained by numerical simulations
of Eq. (\ref{tdgpe}).
Depending on the values of 
the effective chemical potential and the Raman coupling, the system explores 
three different regimes that will be explained in detail in the following 
sections. For values of the Raman coupling smaller than a critical one 
$\Omega_c$ (solid lines) we have found a vortex trapping regime, where winding number states can not
be exchanged between components. 
We have also explored the influence of the 
ratio $g_{\uparrow\downarrow}/g$ by performing 1D numerical 
calculations of the spinor GPE, which are shown by solid lines and open 
symbols in the inset of Fig. \ref{Fig1}. Open circles correspond to our results 
for $g_{\uparrow\downarrow}/g=0.9954$, whereas triangles correspond to  
$g_{\uparrow\downarrow}/g=0.9$. 

It has been demonstrated that in order to produce spin excitations 
\cite{Son2002,Abad2013}, which are 
the relevant ones for phase slips in a spinor condensate, it is necessary to
overcome an energy gap $\Delta$ given by:
\begin{equation}
\Delta=\Omega\,\sqrt{1+\frac{ng}{\Omega}\left(1-\frac{g_{\uparrow\downarrow}}{g} 
\right) } \, ,
 \label{gap}
\end{equation}
where $n$ is the total density.
For the sake of comparison, we have 
complemented the inset of Fig. \ref{Fig1} with the curves given by Eq. 
(\ref{gap}) for the same numerical values of $\Omega_c$:  
dotted ($g_{\uparrow\downarrow}/g=1.0$), dashed 
($g_{\uparrow\downarrow}/g=0.9954$) and dot-dashed 
($g_{\uparrow\downarrow}/g=0.9$) lines. 
The minimal coupling energy $\hbar\Omega_c$ necessary to produce phase slip is 
of the order of $\Delta$. When $g_{\uparrow\downarrow}=g$, phase slips can be 
produced for arbitrarily small values of the coherent coupling. As the ratio 
$g_{\uparrow\downarrow}/g$ decreases the energy cost for producing phase slips 
increases.

Once this gap is overcome, phase slip is possible. The 
system continuously transits from the Non-Coherent 
Quantum Phase Slip (NCQPS) regime at $\Omega \gtrsim \Omega_c$, where vortex 
exchange between components can be observed at rates different from $\Omega$, to the 
Coherent Quantum 
Phase Slip (CQPS) regime at $\Omega\gg\Omega_c$. We identify a process as 
coherent if it evolves without decay and with a well-defined frequency equal to 
the Raman coupling. 

The dynamical phase diagram also depends on the radius of the 
torus. As the ring geometry constitutes a finite system, a zero-point 
kinetic energy $\hbar^2/mR^2$ is introduced. This energy quantum separates 
winding number states and, as a result, 
the degenerate states $|q_1,q_2\rangle$ and $|q_2,q_1\rangle$ are separated 
by a gap from other winding number states. When the radius of the 
torus increases, the zero-point kinetic energy goes to zero and the energy spectrum 
becomes a continuum. The same occurs when the interaction energy is 
very large, because the energy to produce a vortex is negligible in front 
of the chemical potential. As we will show later, CQPS decays or is even 
absent in these cases. The dynamical phase diagram can also exhibit dramatic 
changes in the immiscible case, where vortex states can split due to phase 
separation \cite{GarciaRipoll2002}.

\section{Coherent Quantum Phase Slip}
\label{cqps}

In order to have an analytical insight into the dynamics of the system, 
one can follow the spirit of the two-mode approximation. 
In the CQPS regime, the condensate wavefunction can be written as  \cite{Williams1999a,Williams1999c,Dum1998}:
\begin{eqnarray}
{{\Psi_\uparrow(\vec r, t)}\choose {\Psi_\downarrow(\vec r, t)}}
=\phi_{q_1}(\vec r) {{\psi_{\uparrow,q_1}(t)}\choose 
{\psi_{\downarrow,q_1}(t)}}+\phi_{q_2}(\vec r) 
{{\psi_{\uparrow,q_2}(t)}\choose {\psi_{\downarrow,q_2}(t)}} \,
\label{rotmodes}
\end{eqnarray}
where  $\phi_{q_j}(\vec r)=\phi_{q_j}(\rho)\times e^{i q_j \theta}$, with $j=1,2$, are eigenvectors of both 
the angular momentum operator $\hat{L}_z$, with eigenvalue $\hbar q_j$, and 
the Hamiltonian without Raman coupling, with eigenvalue $\mu_j$. 
This ansatz neglects any contribution from other modes with charges 
different from $q_1$ and $q_2$. As we will see later, our numerical 
results agree with this assumption, 
since the only eigenvectors that significantly contribute to the dynamics are 
those associated to the winding numbers 
imprinted initially onto the wavefunction. 

After substituting Eq. (\ref{rotmodes}) in the GPE 
(\ref{tdgpe}) one gets two decoupled linear Josephson equations for each winding 
number: 
\begin{eqnarray}
 i\hbar\frac{\partial \psi_{\uparrow, q_1}}{\partial t}=\mu_1\psi_{\uparrow,q_1}
 +\frac{\hbar\Omega}{2} \psi_{\downarrow,q_1} \nonumber\\
 i\hbar\frac{\partial \psi_{\downarrow,q_1}}{\partial t}=\mu_1
 \psi_{\downarrow,q_1}+\frac{\hbar\Omega}{2} \psi_{\uparrow,q_1}\,,
 \label{JEq2}
\end{eqnarray}
and 
\begin{eqnarray}
 i\hbar\frac{\partial \psi_{\uparrow, q_2}}{\partial t}=\mu_2\psi_{\uparrow,q_2}
 +\frac{\hbar\Omega}{2} \psi_{\downarrow,q_2} \nonumber\\
 i\hbar\frac{\partial \psi_{\downarrow,q_2}}{\partial t}=\mu_2
 \psi_{\downarrow,q_2}+\frac{\hbar\Omega}{2} \psi_{\uparrow,q_2}\,.
 \label{JEq3}
\end{eqnarray}

The straightforward solution of these linear systems has the eigenvalues $\mu_j\pm\Omega/2$. 
The energy gap between the two levels is $\Omega$, which is the 
driving frequency, then the solution for the condensate wavefunction is: 
\begin{equation}
\hspace{-1cm}
{{\Psi_\uparrow(\vec r, t)}\choose {\Psi_\downarrow(\vec r, t)}}
=\phi_{q_1}(\rho) e^{i q_1 \theta} {{\cos{\frac{\Omega t}{2}}}\choose {-i\sin{\frac{\Omega 
t}{2}}}}+ \phi_{q_2}(\rho) \,e^{-i(\Delta\mu_q t/\hbar-q_2 \theta 
+\varphi)} 
{{i\sin{\frac{\Omega t}{2}}}\choose {-\cos{\frac{\Omega t}{2}}}}\,,
\label{modelsolu}
\end{equation}
and the corresponding densities read:
\begin{equation}
\hspace{-25mm}
 |\Psi_i|^2= \frac{1}{2}\{|\phi_{q_1}|^2 \cos^2{\left(\frac{\Omega 
t}{2}\right)} + |\phi_{q_2}|^2  \sin^2{\left(\frac{\Omega t}{2}\right)} \pm 
|\phi_{q_1}||\phi_{q_2}| \sin{(\Omega t)}\sin{(\Delta \mu_q - \Delta q\, 
\theta+\varphi)}\}\,,
 \label{teoden}
\end{equation}
where  $\Delta q=q_2-q_1$ is the initial winding number imbalance, $\Delta 
\mu_q=\mu_2-\mu_1$ is the associated chemical potential imbalance and $\varphi$ 
is an arbitrary phase.

From Eq. (\ref{modelsolu}) one can get the mean angular momentum imbalance per particle 
$\Delta L_z=(\langle \Psi_\uparrow|L_z|\Psi_\uparrow\rangle-\langle \Psi_\downarrow |L_z|\Psi_\downarrow\rangle)/\hbar N$, 
as a function of time:
\begin{equation}
{\Delta L_z}= \frac{\Delta q}{2}\cos(\Omega t)\,.
\label{meanang}
\end{equation}
This expression predicts that the exchange of vortices oscillates with the 
coherent coupling frequency $\Omega$. As a consequence, a $\pi$-pulse  
exchanges the winding numbers between components, and a $\pi/2$-pulse will 
drive  each component to a quantum superposition of flows with winding numbers 
$q_1$ and $q_2$. 

\subsection{Phase slip between adjacent winding numbers}
\label{singlemode}

\begin{figure}[t!]
\centering
\includegraphics[width=0.7\linewidth, clip=true]{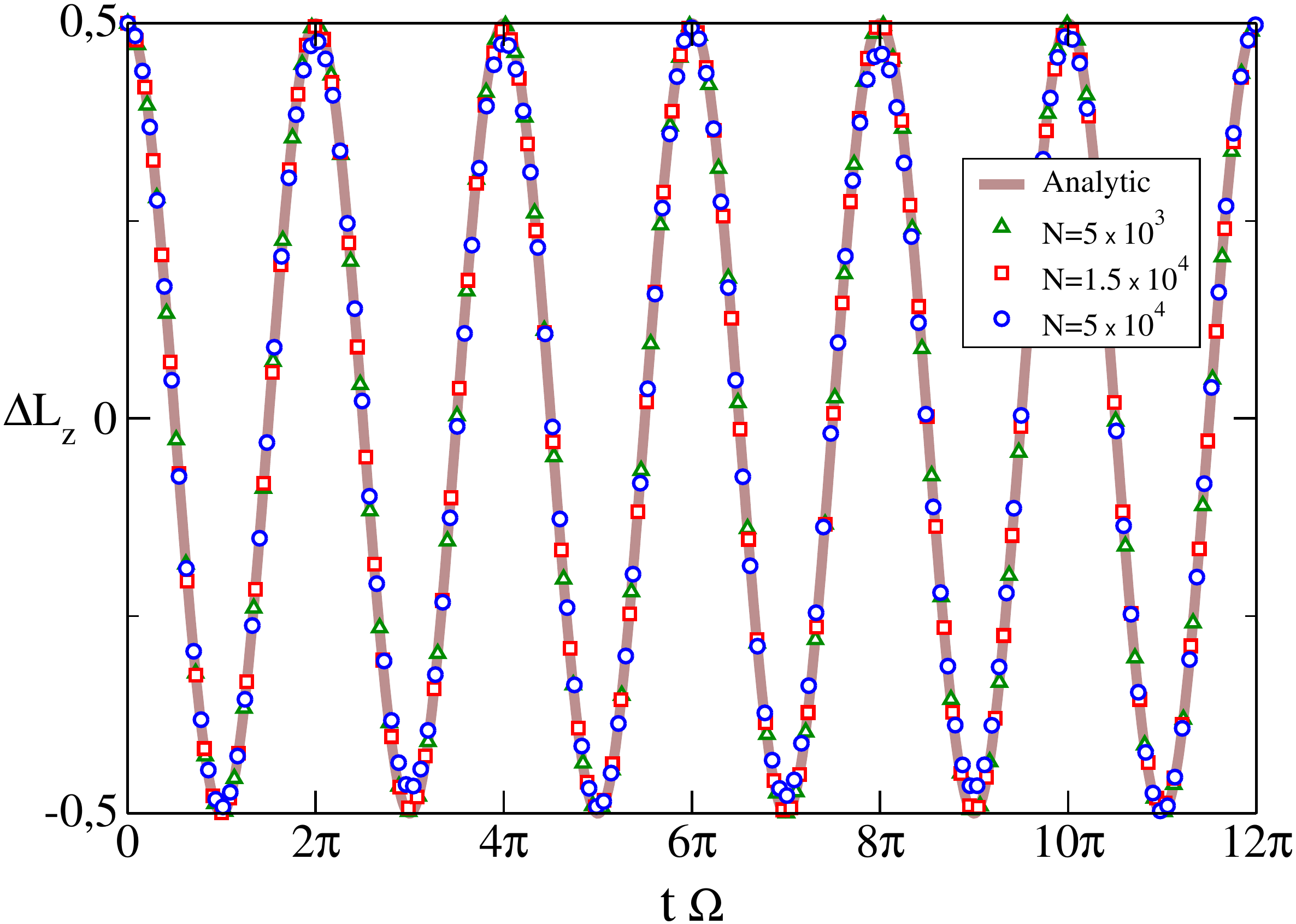}
\caption{Comparison between the mean angular momentum imbalance per particle 
calculated 
by solving the GPE for $N=5\times10^3$ (green 
triangles), 
$N=1.5\times10^4$ (red squares) and $N=5\times10^4$ (blue circles) 
together with the result predicted by Eq.  (\ref{meanang}) (thick brown line). 
The Raman coupling is $\Omega=200\mbox{Hz}$ and the initial state is 
$|q_1=1,q_2=0\rangle$.}
\label{Fig2}
\end{figure} 

Figure \ref{Fig2} shows our numerical results, within the CQPS regime, for the 
mean angular momentum imbalance per particle, obtained 
by solving the GPE (\ref{tdgpe}) in condensates with different number of 
particles and $\Omega=0.16\, \omega_\rho$ (which corresponds to $200$ Hz).
The comparison with the analytical prediction 
given by Eq. (\ref{meanang}) is also shown. As can be seen, the agreement is 
very good. The frequency of the oscillation of 
$\Delta L_z$ is precisely the Raman 
coupling $\Omega$ that coherently connects both spin components. 
The population imbalance  
is initially zero, and remains unaltered during 
the whole simulation, thus one can deduce that the spin 
exchange occurs at pairs even though the population of both 
spin components is not fixed.

\begin{figure}[t!]
\centering
\includegraphics[width=1.0\linewidth, clip=true]{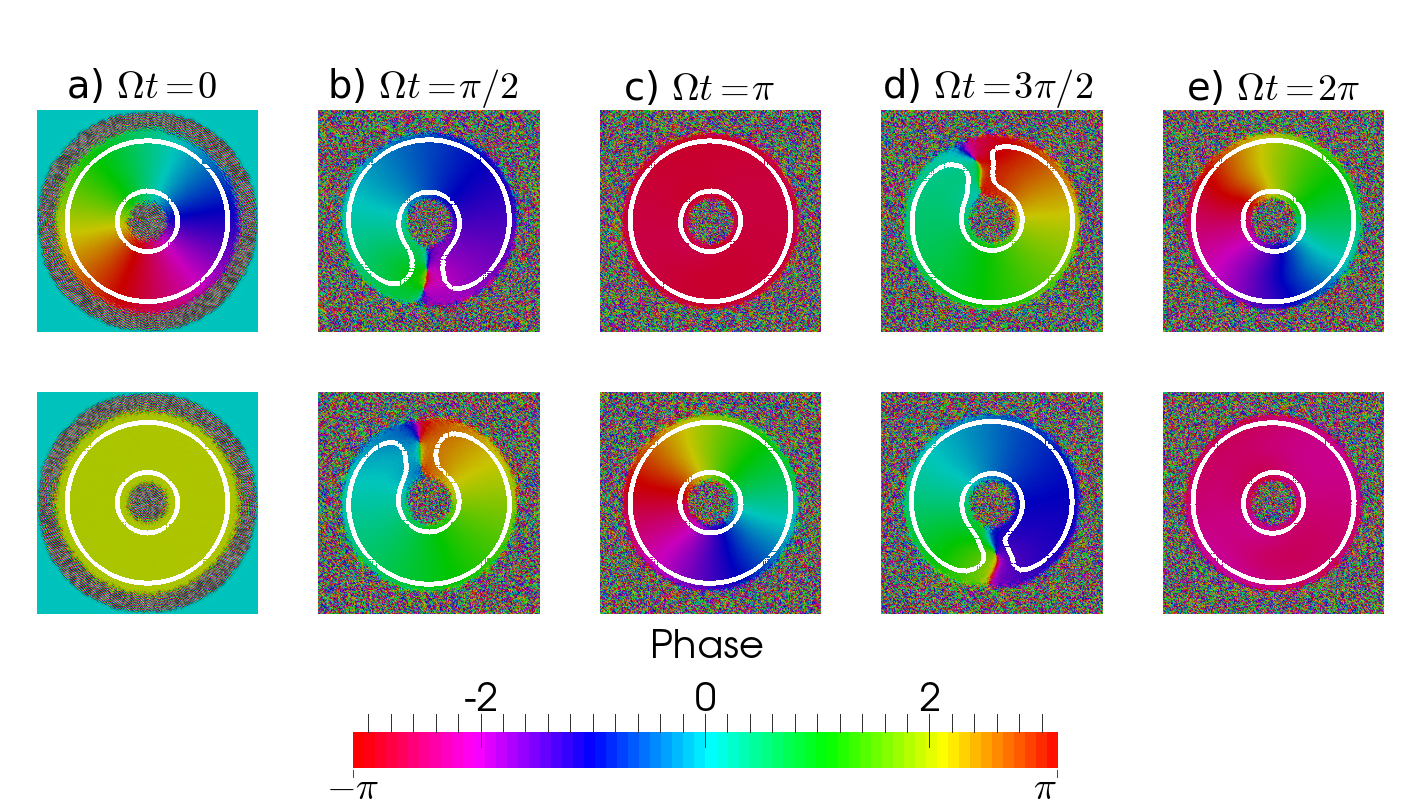}
\caption{Evolution of a condensate with $N=5\times10^4$ atoms as a function of time, 
after imprinting a vortex on the 
$\uparrow$ component (first row). The second row corresponds to the $\downarrow$ 
component.
The value of the Raman coupling is $\Omega=200\,\mbox{Hz}$ and the length of the 
square graphs 
is $30 \mu m$. The white line corresponds to density isocontours at $5\%$ of 
maximum density, whereas 
colours represent the phase. Panels (a-e) display snapshots of the state during 
a Rabi cycle.
This number of particles is in the limit of the CQPS regimes, since $\Delta L_z$ 
deviates in $6\%$ from the 
analytical model.}
\label{Fig3}
\end{figure}

In order to elucidate how the topological structure of the wavefunction changes as a function of time, 
we show in Fig. \ref{Fig3} the dynamical evolution of the density and the local phase 
for a condensate with $N=5\times10^4$ atoms. 
White lines represent density isocontours at $5\%$ of maximum density and colours depict 
the phase. Initially, at $t=0$, we have imprinted a persistent current ($q_1=1$) 
in the $\uparrow$ component while the other is at rest (panel (a)). 
A quarter of period later, $\Omega t=\pi/2$, an azimuthal density node 
\footnote{
This objects should not be confused with dark solitons, since they can also appear in 
the linear case and the associated healing length is, in general, much larger than the one of 
solitary waves. 
}
is formed 
spontaneously in each component, at opposite positions (panel (b)). 
The vortex that 
was inducing the rotation in the initial state 
escapes from the $\uparrow$ component through the density depletion, 
while another vortex crosses the corresponding node in the $\downarrow$ 
component, transferring vorticity from the $\uparrow$ component to the $\downarrow$ component 
(panel (c)). This is the mechanism followed by the coupled system to produce $2\pi$-phase slips. 
After that, in panel (d) the evolution is reversed, 
returning the vorticity to the $\uparrow$ component (panel (e)), and so on.

\begin{figure}[t!]
\centering
\includegraphics[width=0.7\linewidth, clip=true]{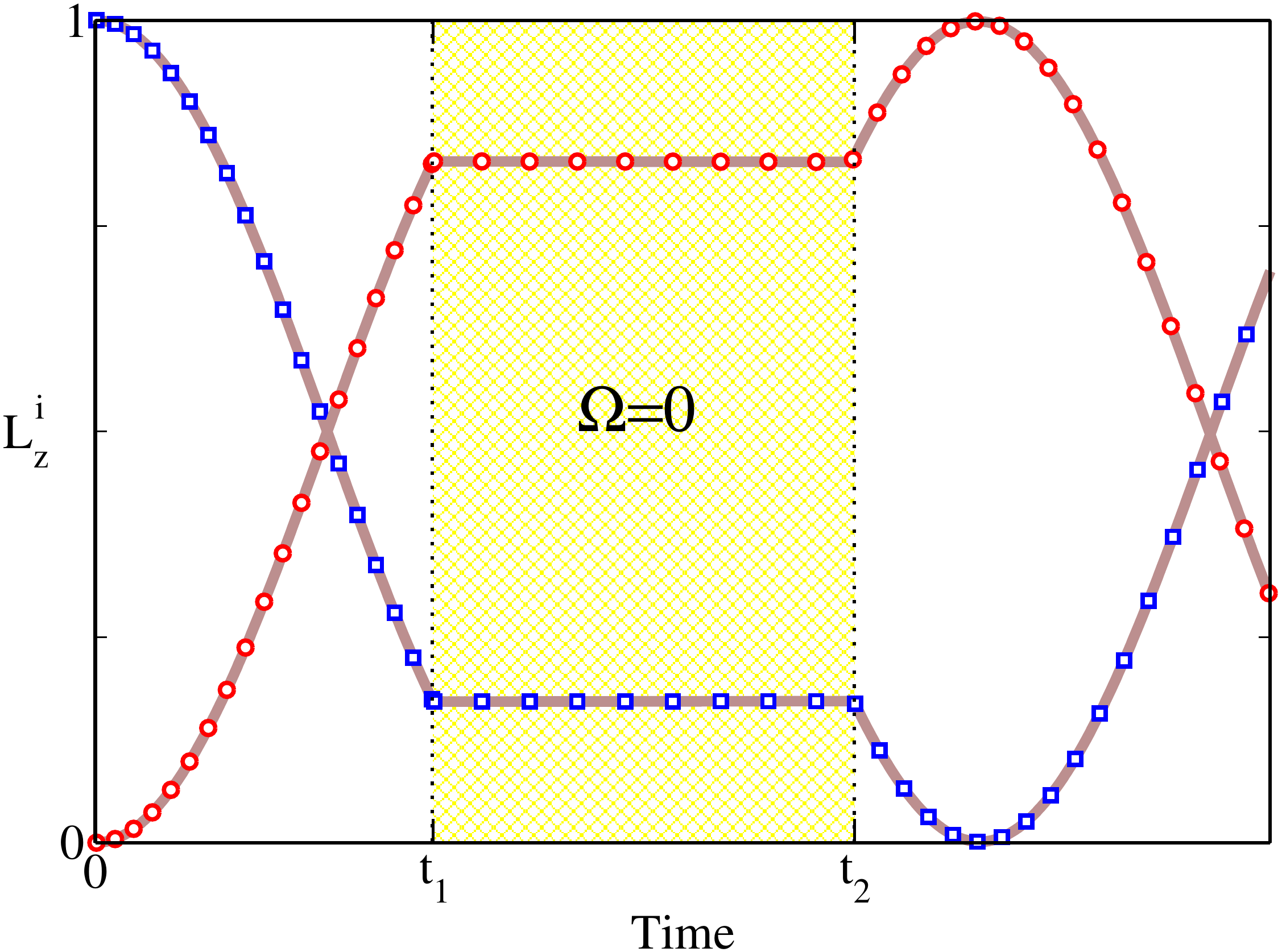}
\caption{Comparison between the analytical model given by Eq. (\ref{meanang}) (thick brown lines) 
and the numerical 
solution of the GPE (open symbols) for the time evolution following the protocol described in the text. 
Blue squares correspond to the mean angular momentum per particle of 
the $\uparrow$ component, and red circles to the $\downarrow$ component. 
The initial state is $|q_1=1$,$q_2=0\rangle$, the number of 
atoms is $N=5\times10^3$ and the Raman coupling is $\Omega=200\,\mbox{Hz}$, except between 
time $t_1 \Omega=14.37/2\pi$ and $t_2 \Omega=32.32/2\pi$, where $\Omega$ is switched off.}
\label{Fig4}
\end{figure}

CQPS in atomtronic circuits allows the system to effectively operate as a qubit.
A quantum mechanical system is a good candidate for qubit manipulation if two 
conditions are fulfilled. First, the system must be considered as an effective 
two-state system, and second, at every time, the qubit must be expressed in a quantum 
superposition of both states. In our system these two states are $|q_1,q_2\rangle$ 
and $|q_2,q_1\rangle$, and as a result the state of the system can be written as 
$\Psi=\alpha |q_1,q_2\rangle+\beta |q_2,q_1\rangle$ at every time. 
From the analytical model, we know that $\alpha=\cos{(\Omega t/2)} \sigma_z$ and 
$\beta=i\sin{(\Omega t/2)}\sigma_z$, where $\sigma_z$ is a Pauli matrix. 
It can be mapped to the most general expression of a qubit 
$\Psi=\cos{(\theta/2)}|0\rangle+\exp{(i\varphi)}\sin{(\theta/2)}|1\rangle$. The mapping is characterized by 
a periodic evolution with a period of $2\pi/\Omega$. 
At half a period, the phases of both components 
are exchanged, and in between, topological defects
appear in the wavefunction in order to drive the phase slip. 
The Raman coupling $\Omega$ is a parameter that can be externally 
manipulated, and its control allows to simulate 
a tunable single-qubit quantum gate for quantum information processes.
In this regime, the system displays two characteristic properties: 
\begin{itemize}
 \item The system 
 behaves as linear despite the non-linearity, since the Rabi frequency is the Raman 
 coupling, independently of the interaction. 
 \item The evolution occurs following  quasi-stationary states. 
\end{itemize}

The two previous remarkable properties can be demonstrated by following this protocol: 
a) Evolve the system with a certain value of $\Omega$ in the CQPS regime, thus the 
angular momentum imbalance will oscillate with frequency $\Omega$. b) At an arbitrary 
time $t_1$, switch the Raman coupling to zero, suppressing the exchange of phase between 
components. c) At another arbitrary time $t_2$, switch on again the Raman coupling to its 
initial value in the process. Figure \ref{Fig4} represents the whole sequence of this 
protocol. At $t_1$ the state gets frozen in a quasi-stationary state rotating at a given 
velocity according to the winding number chemical potential imbalance. Then, at $t_2$ 
the evolution resumes, with exactly the same properties of the system at $t_1$. The 
curves predicted by the analytical model accurately fit in with the solution of the GPE.

\subsection{Phase slip between non-adjacent winding numbers}
\label{multimode}

All the results presented in Sect. \ref{singlemode} are devoted 
to the case of CQPS between adjacent winding number states. However, 
the analytical model is more generic 
and also applies between non-adjacent winding numbers. In this 
section we present the performance of CQPS in the case where the winding 
numbers imprinted onto both wavefunctions differ in more than one unit. 

\begin{figure}[h!]
\centering
\includegraphics[width=0.7\linewidth, clip=true]{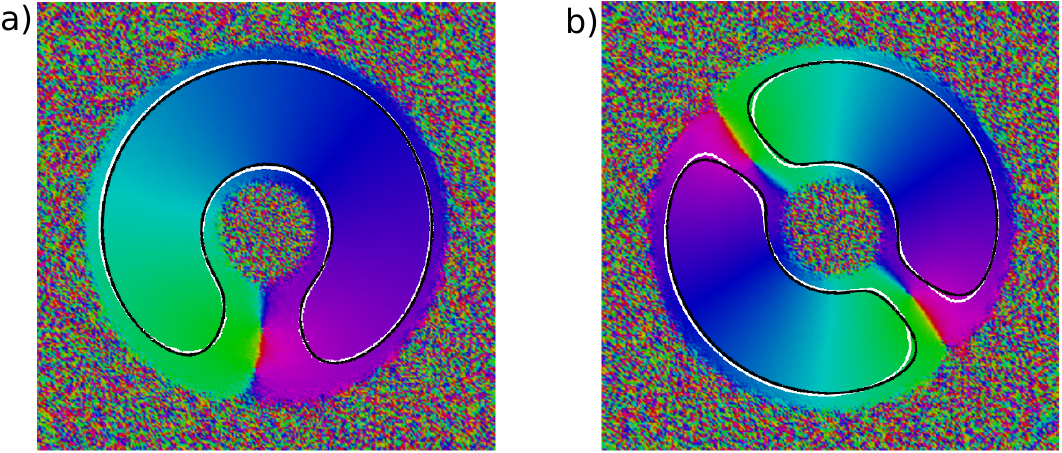}
\caption{Comparison of the numerical results of the GPE with the analytical prediction of 
Eq. (\ref{teoden}), for the wavefunction of the $\uparrow$ component 
at a quarter of a Rabi cycle. The initial state is $|q_1=1,q_2=0\rangle$ 
(panel (a)) and $|q_1=2,q_2=0\rangle$ (panel (b)), the Raman coupling is 
$\Omega=200\,\mbox{Hz}$ and the condensate holds $N=5\times10^4$ atoms. 
White lines correspond to density isocontours at $4\%$ of maximum density 
and colours to the phase, both of them obtained numerically. Black lines 
correspond to density isocontours at $4\%$ of maximum density predicted 
by the model, assuming the initial density in the Thomas-Fermi limit, 
given by Eq. (\ref{TF}).}
\label{Fig5}
\end{figure} 

\begin{figure}[b!]
\centering
\includegraphics[width=0.7\linewidth, clip=true]{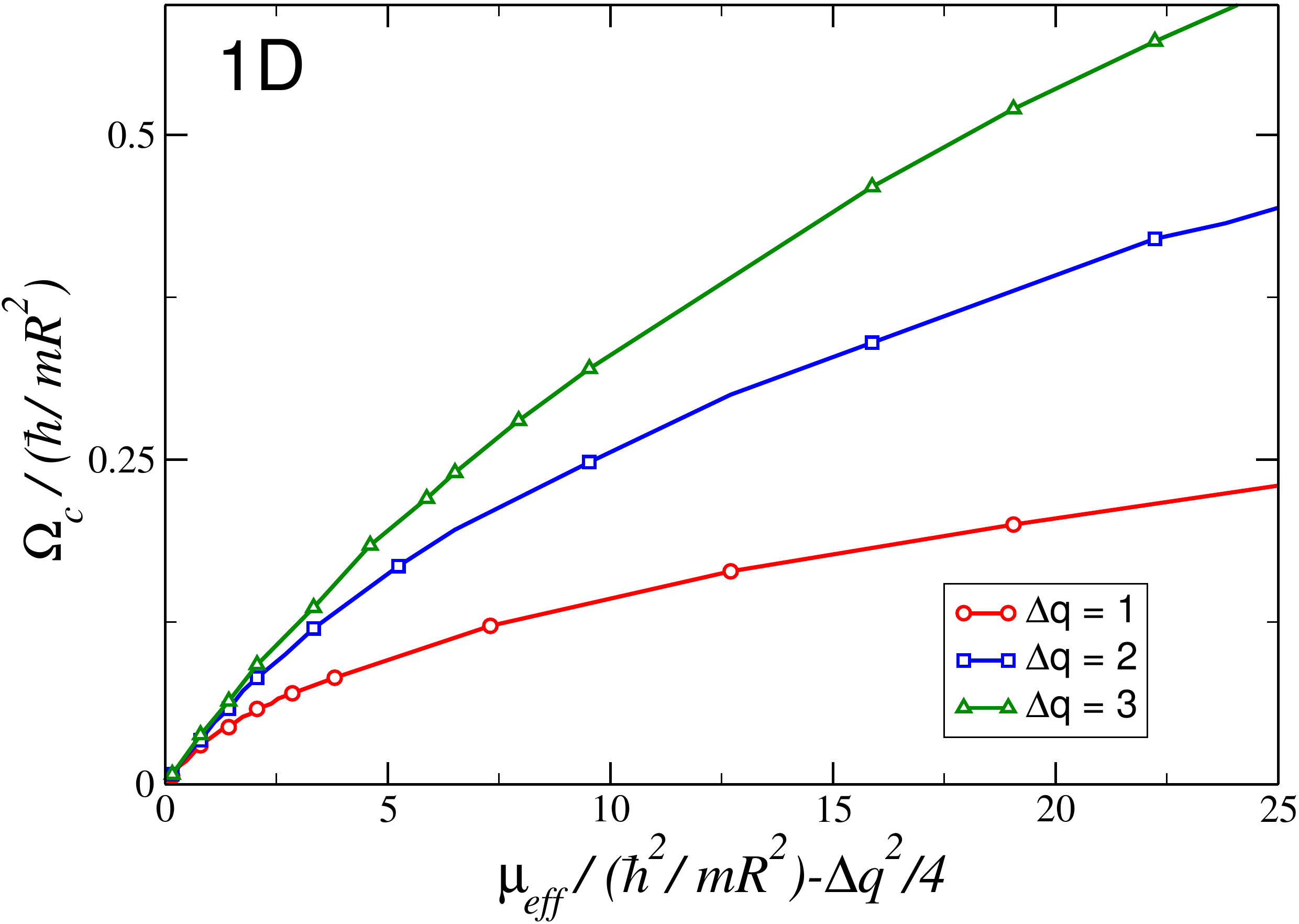}
\caption{Critical coupling $\Omega_c$ as a function of the effective chemical potential 
$\mu_{\rm eff}$ for different values of the winding number imbalance $\Delta q=1$ (red circles), 
$\Delta q=2$ (blue squares) and $\Delta q=3$ (green triangles), with $q_2=0$. The 
results have been obtained by solving numerically the 1D-GPE with 
$g_{\uparrow\downarrow}/g=0.9954$. 
}
\label{Fig6}
\end{figure} 

\begin{figure}[h!]
\centering
\includegraphics[width=0.99\linewidth, clip=true]{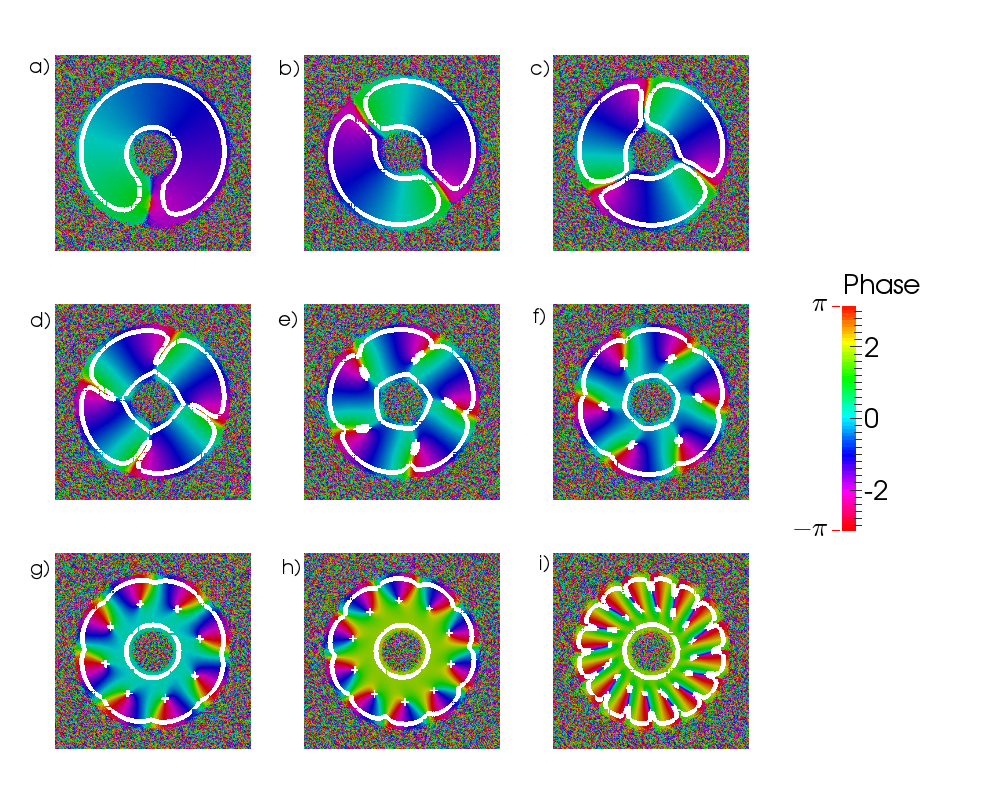}
\caption{Density isocontours at $5\%$ of maximum density and phase (colour) 
of the $\uparrow$ component at a quarter of a Rabi cycle for a condensate of $N=5\times10^4$ 
atoms, $\Omega=200\,\mbox{Hz}$ and different values of the initial angular 
momentum imbalance, with the $\downarrow$ component first at rest. a) $q_1=1$, b) $q_1=2$, 
c) $q_1=3$, d) $q_1=4$, e) $q_1=5$, f) $q_1=6$, g) $q_1=8$, h) $q_1=10$ and i) $q_1=16$.
Only the cases of the first row do exhibit CQPS.
}
\label{Fig7}
\end{figure} 

We have shown that in order to change the winding number in one unit, a 
$2\pi$-phase slip event involving the formation of an azimuthal density node 
has to occur. Therefore, to drive 
each component from winding number $q_1$ to $q_2$ (and viceversa), 
multiple number of such nodes ($|q_1-q_2|$) must appear simultaneously. 
Multiple $2\pi$-phase slip can not occur through a sequence of single $2\pi$-phase 
slip events, since other states with winding number different from $q_1$ 
and $q_2$ would contribute, as described by the ansatz (\ref{rotmodes}). 
This fact can be seen in Fig. \ref{Fig5}, where we compare the numerical 
results of the GPE (white isocontours at $4\%$ of maximum density and colours for the phase), 
with the analytical prediction for the same density isocontour given by Eq. (\ref{teoden}) 
(black line), assuming that the system is in the 
Thomas-Fermi limit and Eq. (\ref{TF}) applies, for the initial state $|q_1=1,q_2=0\rangle$ 
(panel (a)) and $|q_1=2,q_2=0\rangle$ (panel (b)). The agreement is again excellent. 

Fig. \ref{Fig6} represents $\Omega_c$, which  
fixes the critical value of the Raman coupling that allows phase slip events, 
as a function of the effective chemical potential, for different 
initial winding number imbalances $\Delta q=1,2,3$ (red circles, blue 
squares and green triangles, respectively), after solving the 1D-GPE with  
$g_{\uparrow\downarrow}/g=0.9954$. 
$\Omega_c$ increases with $\mu_{\rm eff}$, but this increasing is faster for 
larger initial $\Delta q$. The azimuthal nodes 
that the system has to generate in order to produce phase slips possess 
more energy, and the strength of the coherent coupling has to be larger to 
overcome higher energy barriers associated to smaller characteristic lengths. 

We have studied the dynamics of the system for different initial winding number imbalance. 
Figure \ref{Fig7} shows the density (white isocontours)  
and the phase (colour) of the $\uparrow$ component at a quarter of a Rabi cycle, 
for different values of the initial winding number $q_1$ (with $q_2=0$) 
imprinted onto condensates with $N=5\times10^4$ atoms. One can observe that 
density nodes appear equispaced forming an ordered pattern. 
As the initial angular momentum imbalance increases, the 
characteristic length scale associated to the azimuthal density nodes decreases as predicted by Eq. (\ref{teoden}), 
and becomes closer to that of solitary waves as dark solitons or solitonic vortices. 
To generate such objects additional winding number modes have to be excited, and therefore, 
the CQPS process decays after few cycles. Then, the 
ansatz (\ref{rotmodes}) is no longer valid. That is why the second and the third row 
of Fig. \ref{Fig7} will not exhibit CQPS (they belong to the NCQPS regime), and 
only the cases of the first row will display this phenomenon. Each solitonic 
vortex that mediates the phase slip 
in the cases falling in the NCQPS regime contributes to the total mean angular 
momentum of each component according to Eq. (\ref{off-centred-vortex}).

\section{Other dynamical regimes}
\label{Other}

As was shown in Fig. \ref{Fig1}, CQPS can not be found in the whole parameter 
space of the system, given that $g\neq g_{\uparrow\downarrow}$. Our 
numerical results point to the fact that CQPS exists as far as the ansatz 
(\ref{rotmodes}) is valid, and this occurs for high values of the coherent 
coupling in comparison with $\Omega_c$. Below this critical value, we have 
found a dynamical regime where there is no vortex exchange between spin 
components. This fact is due to the presence of an energy barrier  
between winding number states $|q_1$,$q_2\rangle$ and $|q_2$,$q_1\rangle$, that 
prevents phase slips. Although such a barrier is of nonlinear nature, 
contrary to scalar condensates, it is not due to 
the existence of solitonic states on the total density of the system 
\cite{MunozMateo2015}. As mentioned before, the excitation of spin modes has 
been demonstrated to play a key role in spinor condensates coupled by density, 
and seems to be also relevant in this case. Such excitations are separated from 
the winding number states by the energy gap given in Eq. (\ref{gap}), which 
must be overcome in order to produce phase slips. If this critical energy 
can not be transferred between components by the coherent coupling, vortices 
will be trapped, and the mean angular momentum per component will not oscillate 
around the value $(q_1+q_2)/2$.
\begin{figure}[htb]
\centering
\includegraphics[width=0.95\linewidth, clip=true]{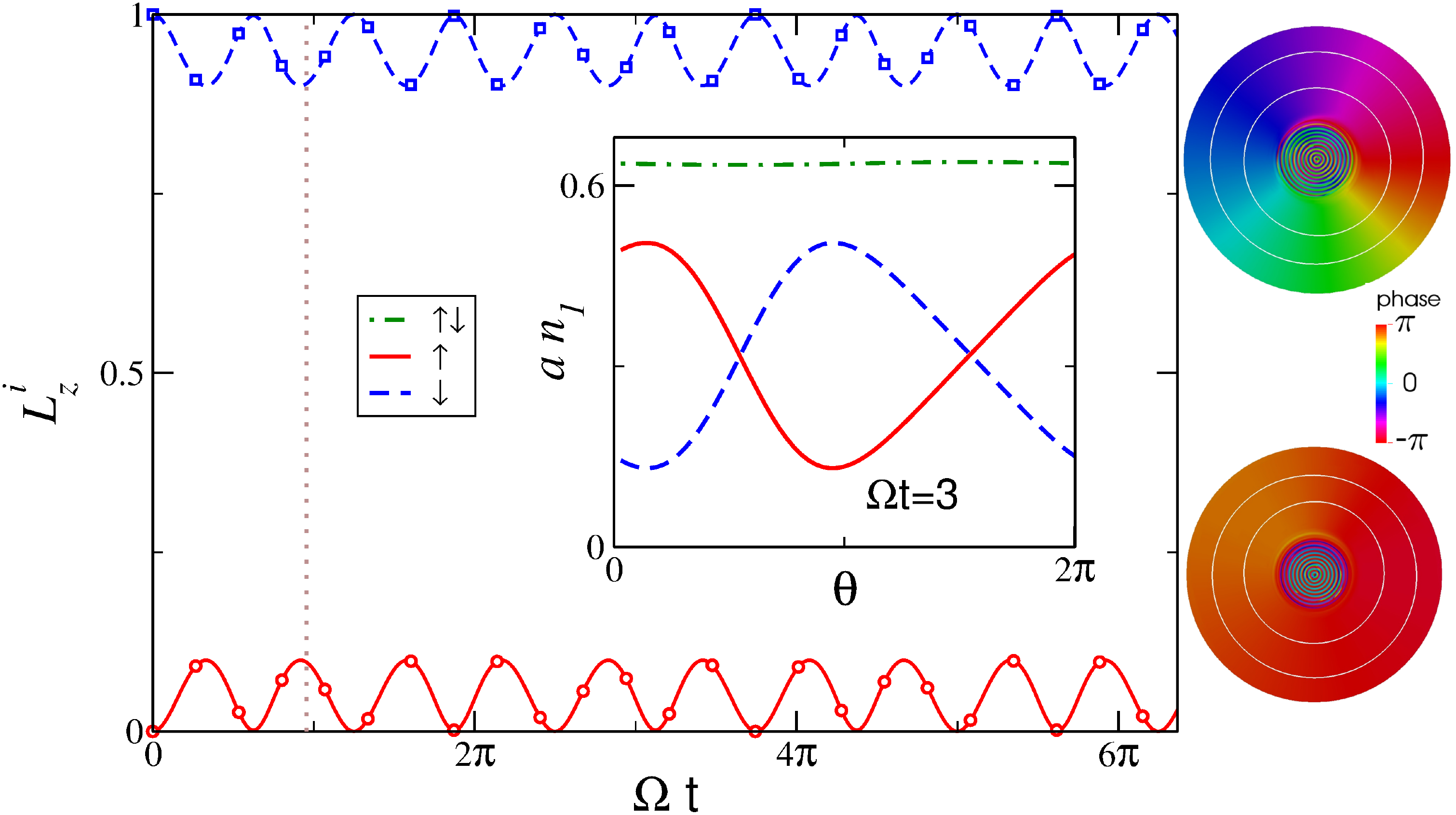}
\caption{Mean angular momentum per particle of the $\uparrow$ (solid 
red line with circles) and $\downarrow$ (dashed blue line with squares) 
component, after solving the 2D-GPE. In the inset, the azimuthal density 
$a\,n_1=(a/R) \int |\phi (\rho, \theta)|^2 \rho d\rho$ of the $\uparrow$ (solid red line), $\downarrow$ (dashed blue 
line) component, and the sum of both (dot-dashed green line), at $\Omega t=3$ 
(indicated by the dotted vertical line). 
At the right side of the plot, phase pattern of the $\uparrow$ (top) and 
$\downarrow$ (bottom) component is represented by colours, and the isocontours at 5\% of maximum 
density by the white lines. The effective chemical potential is 
$\mu_{\rm eff}=\hbar\omega_\rho$ and the Raman coupling is 
$\Omega=2\times10^{-3}\omega_\rho$.}
\label{Fig8}
\end{figure}

Fig. \ref{Fig8} shows a typical case representative of the trapping regime. It 
corresponds to the point A indicated in Fig \ref{Fig1}, for the coupling 
$\Omega=2\times10^{-3}\omega_\rho$. As can be seen, the mean angular momentum 
of each component oscillates near the initial value, and the corresponding 
densities (shown in the inset after integration along the transverse section 
of the torus) present variations without nodal points. Although the 
interaction between components translate into currents inside each component 
(see the phase maps on the right of the figure), they are not enough to 
drive phase slips. Finally it is worth to note, that during all the 
time evolution the total density remains approximately constant along the torus.

\begin{figure}[htb]
\centering
\includegraphics[width=0.95\linewidth, clip=true]{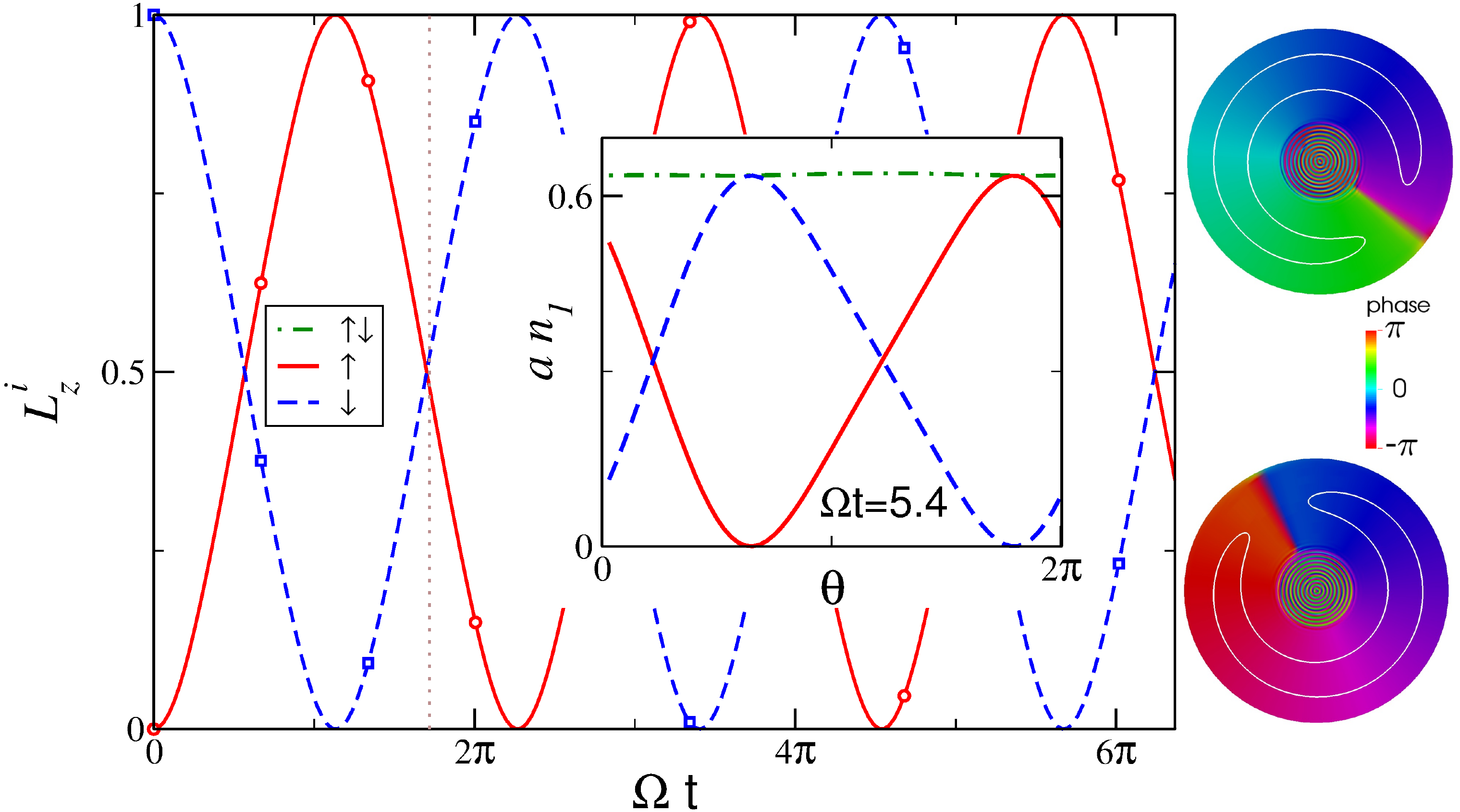}
\caption{Same as Fig. \ref{Fig8} for $\Omega=5\times10^{-3}\omega_\rho$ at $\Omega t=5.4$.
}
\label{Fig9}
\end{figure}

When the Raman coupling takes intermediate values, $\Omega\gtrsim\Omega_c$, 
stable CQPS will not manifest 
in the dynamics, and the system enters the NCQPS regime. In this case, 
the coherent coupling is large enough to produce phase slip events that 
exchange the winding number between spin components. However, the time 
frequency of these events is lower than $\Omega$. This 
features are reflected in the case displayed in Fig. \ref{Fig9}, corresponding 
to the point B of Fig. \ref{Fig1}. Now the spin densities can show nodal points 
leading to phase slips, whereas the total density remains constant. Contrary to 
the CQPS case, the position of such nodal points for both components are not 
located at diametrically opposed positions. As the coherent coupling strength 
increases the frequency for phase slips approaches to $\Omega$. 

\begin{figure}[h!]
\centering
\includegraphics[width=0.55\linewidth, clip=true]{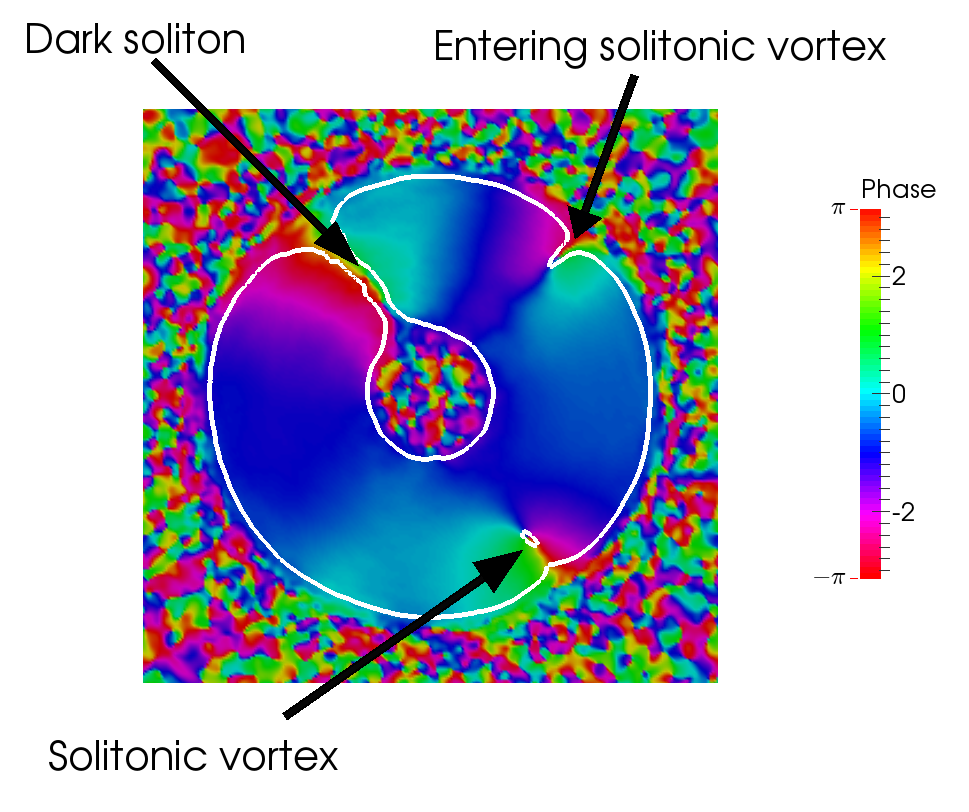}
\caption{Density isocontour at $5\%$ of maximum density and phase (colour) of 
the $\uparrow$ component of a condensate of $N=5\times10^4$ atoms and 
$\Omega=60\mbox{Hz}$ 
after $2\,s$ of evolution, with the state $|q_1=1$,$q_2=0\rangle$ as the 
initial 
state. Two solitonic vortices and a dark soliton appear in the wavefunction. }
\label{Fig10}
\end{figure} 

In addition, as the chemical 
potential increases, the excitation of solitary waves (see panels (d)-(i) in 
Fig. \ref{Fig7}) are responsible for the damping of the exchange of 
angular momentum between components. As a consequence, the system can deviate 
from the quasi-stationary path and explore other regions of the phase space. 
Different topological objects are generated, and then, the long-time 
dynamics will bring the condensate to an out-of-equilibrium 
quantum gas. Notice that in this regime many angular momentum modes are excited 
and the two-mode approach (\ref{rotmodes}) is no longer valid. 
In Fig. \ref{Fig10} we show a characteristic snapshot of the density and the 
phase of the $\uparrow$ 
component of a condensate with $N=5\times10^4$ atoms and Raman coupling 
$\Omega=60\,\mbox{Hz}$ after $2\,$s of evolution from the initial state 
$|q_1=1$,$q_2=0\rangle$. 
The white line traces the density isocontour at $5\%$ of maximum density 
and the colours display the phase. In the figure one can also see that although 
the initial angular momentum imbalance is $\Delta q=1$, several kinds of 
solitary waves (a dark soliton and two solitonic vortices) are excited in the 
condensate in order to try to drive a single $2\pi$-phase slip. One can 
see that the appearance of solitonic vortices do not accomplish with the 
azimuthal dependence of the density predicted by Eq. (\ref{teoden}), and as a 
consequence, it is not compatible with CQPS.

\section{Summary and conclusions}
\label{conclusion}

In the present work we have proposed an atomic analogue of the Mooij-Harmans qubit 
that displays Coherent Quantum Phase Slip. Two-component condensates loaded on 
toroidal atomtronic circuits can display phase slips by virtue of the coherent 
coupling. When a vortex pattern phase is imprinted onto each component with 
different winding number, the system evolves through quasi-stationary states that are a superposition 
of both winding number states. The two components exchange vortices by phase 
slip events modulated by the coupling, in such a way that the mean angular momentum 
imbalance oscillates with the Raman frequency. 

We have identified the different dynamical regimes of the system as a function 
of the coherent coupling and the effective chemical potential. In particular, we 
have focused on the dynamical phase corresponding to CQPS, where the 
system behaves effectively as linear despite the non-linearity. For this regime, 
we have mapped the dynamics of the coupled system onto linear Josephson equations by using an 
ansatz composed of two winding number modes per component. The whole dynamics, 
and specifically the results obtained for the mean angular momentum imbalance 
and the density, are very accurately reproduced by our analytical model. This 
model predicts that CQPS needs phase slip events to occur through azimuthal density nodes, 
otherwise coherence would be destroyed. 
Our numerical results obtained by solving the time-dependent GPE confirm these 
predictions. 

We would like to point out the experimental feasibility of this system, 
since we have used values for the physical parameters currently available in 
laboratories. $^{87}\mbox{Rb}$ is a good candidate to perform this qubit, mainly, 
for the closeness of the scattering lengths. Toroidal condensates that have been recently 
obtained have a diameter of the order of 10-20 $\mu$m, 
while ours is 15 $\mu$m. The Raman coupling does not present any  
strong limitation for its value although commonly, it ranges from tens of Hz, up to few 
hundreds. Besides, phase imprinting techniques have improved in the last decade 
and the individual manipulation of a single component of a spinor BEC is 
possible nowadays.

The qubit we have proposed points to multiple possibilities in the field of 
cold atoms. The control of the coherent coupling permits to freeze the system 
in quasi-stationary persistent current states with non-quantized angular momentum 
characteristic 
of linear superposition of quantum states. Since both current states are entangled, 
one can manipulate the quantum superposition of both flow states, 
performing as a good quantum computer gate, and offering paths for improvements 
in quantum information processing. 
The theoretical analysis discussing the role of non-linear objects, as relative 
phase domain walls and dark solitons, are out of the scope of the present article and 
will be addressed in the future.

\ack
We  acknowledge  financial  support  from  the  Spanish  MINECO (FIS2011-28617-C02-01 and 
FIS2014-52285-C2-1-P) 
and the European Regional development Fund, Generalitat de Catalunya Grant No.  2014 
SGR 401.  A.G. is supported by Spanish MECD fellowship FPU13/02106. We thank useful 
discussions with G. Baym.

\section*{References}

\bibliographystyle{iopart-num}
\bibliography{bibtex}

\end{document}